# Scalable Reshaping of Diamond Particles via Programmable Nanosculpting

*Tongtong Zhang,*[1, 2, ‡] *Fuqiang Sun,*[3, ‡] *Yaorong Wang,*[4] *Yingchi Li,*[1] *Jing Wang,*[5] *Zhongqiang Wang,*[6] *Kwai Hei Li,*[7] *Ye Zhu,*[8] *Qi Wang,*[6] *Lei Shao,*[5] *Ngai Wong,*[1] *Dangyuan Lei,*[4] *Yuan Lin,*[2, 3, *] *and Zhiqin Chu*[1, 2, 9, *]

[1] Department of Electrical and Electronic Engineering, The University of Hong Kong, Pokfulam Road, Hong Kong, China

[2] Advanced Biomedical Instrumentation Centre, Hong Kong Science Park, Shatin, New Territories, Hong Kong, China

[3] Department of Mechanical Engineering, The University of Hong Kong, Pokfulam Road, Hong Kong, China

[4] Department of Material Science and Engineering, City University of Hong Kong, Kowloon, Hong Kong, China.

[5] State Key Laboratory of Optoelectronic Materials and Technologies, Guangdong Province Key Laboratory of Display Material and Technology, School of Electronics and Information Technology, Sun Yat-sen University, Guangzhou, China

[6] Dongguan Institute of Opto-Electronics, Peking University, Dongguan, China

[7] School of Microelectronics, Southern University of Science and Technology, Shenzhen, China

[8] Department of Applied Physics, Research Institute for Smart Energy, The Hong Kong Polytechnic University, Hung Hom, Hong Kong, China

[9] School of Biomedical Sciences, The University of Hong Kong, Hong Kong, China

**ABSTRACT:** Diamond particles have many interesting properties and possible applications. However, producing diamond particles with well-defined shapes at scale is challenging because diamonds are chemically inert and extremely hard. Here, we show air oxidation, a routine method for purifying diamonds, can be used to precisely shape diamond particles at scale. By exploiting the distinct reactivities of different crystal facets and defects inside the diamond, layer-by-layer outward-to-inward and inward-to-outward oxidation produced diverse diamond shapes including sphere, twisted surface, pyramidal islands, inverted pyramids, nano-flowers, and hollow polygons. The nanosculpted diamonds had more and finer features that enabled them to outperform the original raw diamonds in various applications. Using experimental observations and Monte Carlo simulations, we built a shape library that guides the design and fabrication of diamond particles with well-defined shapes and functional value. Our study presents a simple, economical and scalable way to produce shape-customized diamonds for various photonics, catalysis, quantum and information technology applications.

**INTRODUCTION**

Diamond particles have remarkable attributes including unparalleled hardness, superior thermal conductivity, wide band gap, and biocompatibility.[1-3] At the micro and nano scale, the particle geometry defines their optical properties,[4] color center performance,[5] catalytic ability,[6] and their interactions with biological cells.[7] Finding ways to control the shape of the diamond particles at the nano- and micro-scale is thus, key to further unlocking their potential as a material.

Different methods to manipulate and shape materials at the nano or atomic level exist. For example, controlled crystallographic growth,[8] self-assembly,[9-11] and 3D printing[12] have been used to produce complex structures such as colloidal quasicrystals,[13] micron-size DNA origami slats,[14-16] and multi-shape metals[17-19] and metal-organic framework (MOF) particles.[20-22] Custom-shaped diamonds for different applications have also been produced using reactive ion etching (RIE) and electron-beam lithography (EBL).[23-26] While precise, they are not scalable when applied to realistic diamond particles with irregular shapes and internal crystal structures. They are also expensive, slow, and have low yield. Bottom-up synthetic methods such as chemical vapor deposition (CVD) are also not ideal because diamond growth is unpredictable and requires extreme pressure and/or temperature.[27] 3D printing, whilst popular for fabricating different 3D shaped materials,[12] is not precise enough[28] for turning nano- or micro-diamonds into the desired and relevant shapes for photonics and quantum applications.

Air oxidation is a promising method to sculpt diamonds. The method uses atmospheric oxygen to oxidise carbon atoms into carbon dioxide at elevated temperatures. Such thermal oxidation methods have been used to control the surface chemistry[29] and size[30] of nanodiamond powders and remove non-diamond impurities.[31] Given diamond particles have diverse crystal facets[32] and internal defects[33] that exhibit varying levels of reactivity,[25, 34] we posit that air oxidation, which will selectively oxidize the most reactive carbon atoms first, could be a straightforward and simple method for large scale diamond shape engineering. By removing carbon atoms from both the outside and inside of the particle based on the different reactivities of the crystal facets and internal defects, the diamond particles can, in principle, be sculpted into a desired shape. Guided by Monte Carlo simulations, this technique allows for precise control over the sculpting parameters (Figure 1a). The intricacies of the sculpting process are meticulously executed through selective oxidation that progresses from the outer surfaces of the crystal facets to their inner depths, as well as from the innermost regions of the defects outward (Figure 1b). Because multiple diamond particles can be subjected to air oxidation at the same time, the method is scalable. Air oxidation is also easy to implement because it needs only standard equipment and mild experimental conditions.

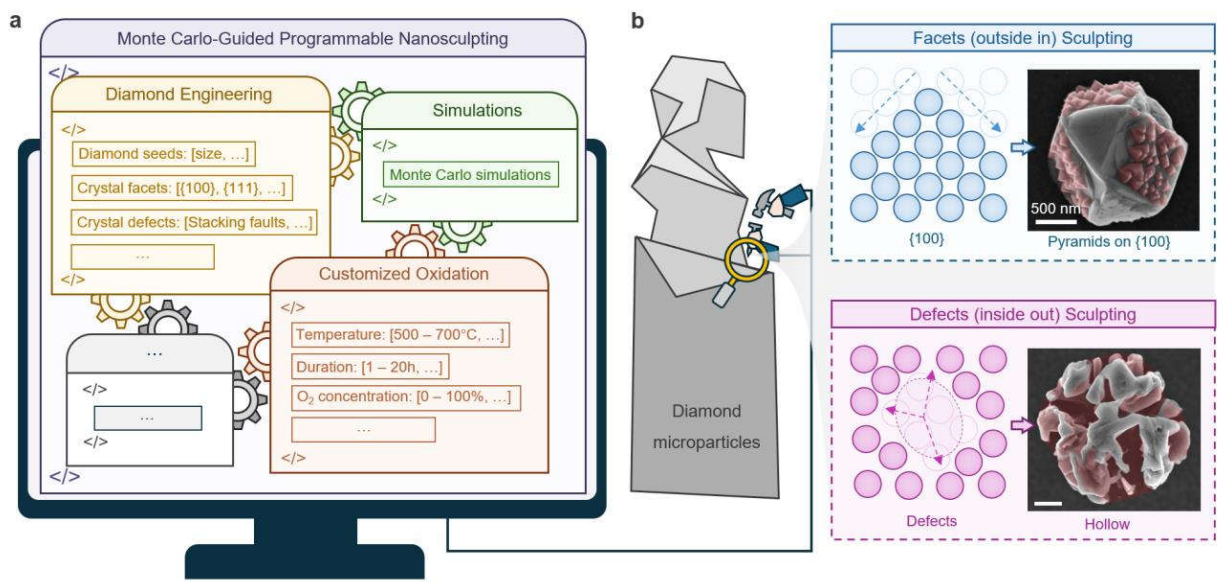

**Figure 1.** Large scale, programmable nanosculpting of diamond particles. Schematic illustrating how large numbers of diamond particles could be sculpted into the desired shapes at once using air oxidation. (a) Diamond seed size, crystal facets, shapes, and defects are considered when using

Monte Carlo simulation to obtain the optimum oxidation parameters such as temperature, time, and oxygen concentration. (b) Nanosculpting of diamond particles exploits the difference in reactivity between crystal facets and defects in the particle. Optimized parameters enable oxidation (sculpting) to progress selectively with precision between the facets and defects, e.g., yielding pyramidal islands on {100} facets and hollow structures (the sculpted parts are pseudo-colored in red).

**RESULTS AND DISCUSSION**

**Air Oxidation Yields Diamond Particles with Distinct Shapes.**

To test our hypothesis, we produced diamond microparticles on a two-inch silicon wafer (Figure 2a) through CVD using salt-assisted air-oxidized nanodiamonds as seeds.[35] These diamond microparticles, which are all distinct from one another, have been shown to be excellent physically unclonable functions (PUFs)[36] – unique identifiers used in high-security applications.[37] We oxidized these microparticles in air at different temperatures (from 500 to 700 °C) for various durations (from 1 to 20 hours) and characterized them.

Scanning electron microscopy (SEM) images show oxidation transformed the raw diamonds into hollow particles (Figure 2b). Upon closer examination, we found the oxidized diamond particles exhibit distinct shapes such as sphere, twisted surface, pyramidal islands, inverted pyramid, flower, and hollow polygons (Figure 2c). Moreover, the original raw diamond particles mainly have two types of shapes: regular and irregular. The regular shaped particles display well-defined facets and edges whereas the irregular diamonds have rough and random surface topologies. These differences suggest that the particles possess distinct crystal and internal structures. The different facets and internal defects in the original raw diamond particle were likely etched at varying rates to produce these distinct morphologies.

We found that besides the initial state of the microparticles, the oxidation temperature and duration also affected the final shape of the diamond. Through a series of experiments, we constructed a phase diagram (or sculpting window) to summarize the sculpting parameters for each diamond shape (Figure 2d and S1). In general, small pyramidal structures tend to form on the outer surface of the diamond particle at low oxidation temperatures while more complicated shapes like flowers and hollow particles are produced when the particles are oxidized at temperatures above 600 °C for a short (5~10 hours) duration. Longer oxidation times etched away the diamond atoms, leaving only a skeleton.

Cross-sectional bright-field (BF) transmission electron microscopy (TEM) image of the hollow diamond particle reveals the presence of holes within the structure (Figure 2e, left panel). A 3D model of the hollow particle constructed using SEM images and computer vision algorithms also displays a clear cavity (Figure 2e, middle and right panels; Movie S1). Further, the oxidized hollow diamond particles retained their crystal structure without any apparent impurities or damage. The X-ray diffraction (XRD) (Figure 2f) and Raman (Figure 2g) spectra of both the hollow and raw diamond particles are identical. SEM images of the same region on the Si wafer before and after air oxidation show that 100% of the particles were uniformly reshaped (Figure S2 and S3), indicating that air oxidation can produce large numbers of consistently shaped diamond particles simultaneously.

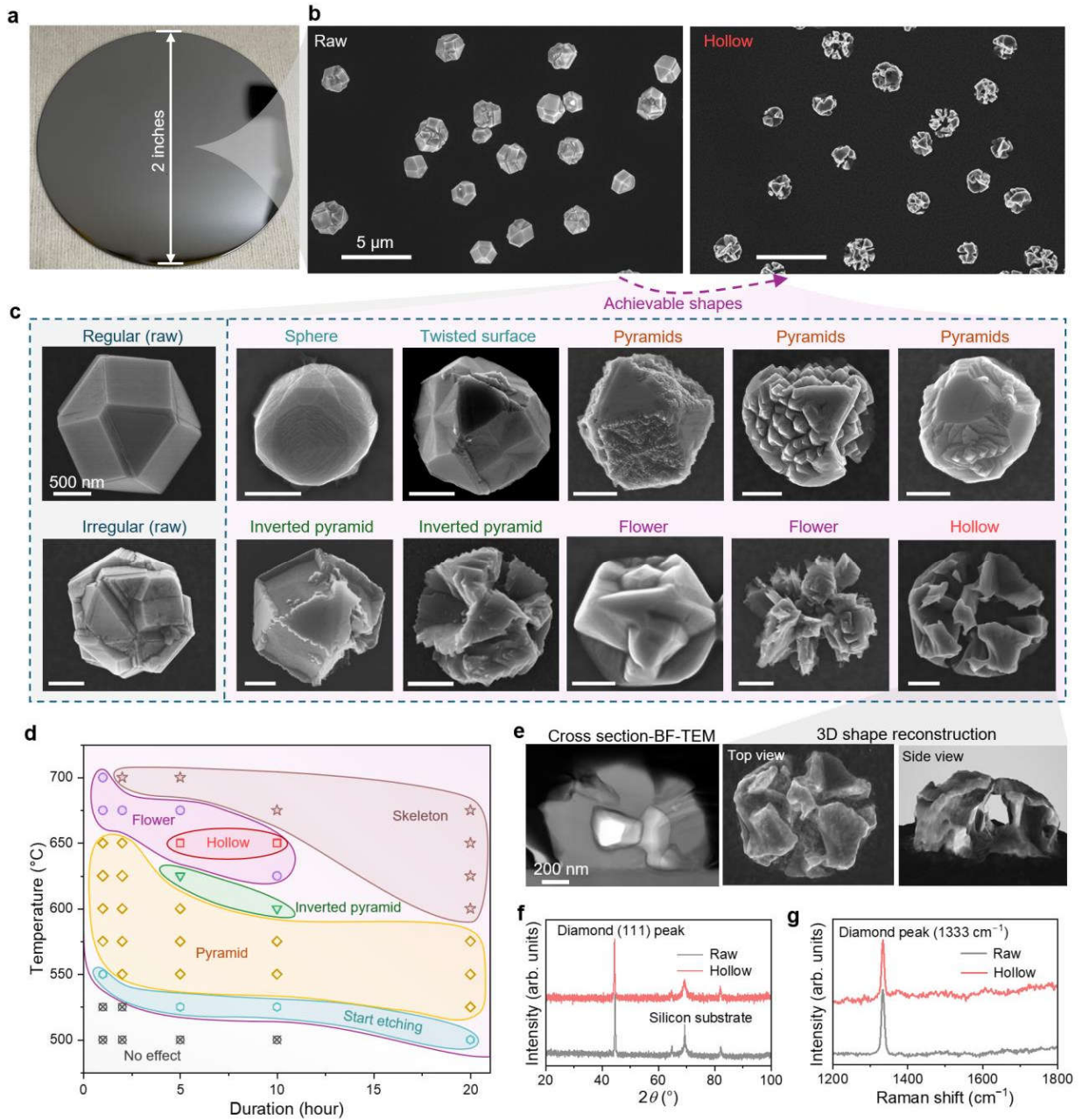

**Figure 2.** Air oxidation produces diamond microparticles with distinct shapes at scale. (a) Photograph of a 2-inch Si wafer with CVD-grown diamond particles illustrates air oxidation is capable of inch-scale nanosculpting. (b) SEM images of raw and sculpted hollow diamond microparticles. (c) SEM images show the original raw diamond particles with regular and irregular shapes, and the nanosculpted diamond particles can have sphere, twisted surface, pyramidal,

inverted pyramidal, flower, and hollow shapes. (d) Phase diagram showing the sculpting parameters (temperature and duration) for each type of diamond shape. (e) Both cross-sectional BF-TEM image (left panel) and the top- (middle panel) and side-view (right panel) of a 3D-reconstructed model of the hollow particle in (b) illustrate distinct holes exist inside the hollow particle. The identical (f) XRD and (g) Raman spectra of raw (grey) and hollow (red) diamond particles show the oxidized hollow diamond particles maintained their diamond crystal structure without any detectable impurities and/or damage.

**Protruding Shapes Are from Etching Crystal Facets Outside In.**

Intrigued by our observations, we sought to understand how the shape transitions occurred. Examining the oxidized diamond particles in Figure 2c, we found that the pyramidal structures formed only on the square-shaped {100} crystal facets but not on the triangular-like {111} crystal facets. Such a selective formation suggests that different crystal facets in the diamond may have distinct oxidation behaviors. To verify this, we monitored the shape evolution of a regular shaped raw diamond microparticle bearing distinct {111} and {100} crystal facets (Figure 3a).

The {100} facets of the raw diamond particle are highly reactive. An initial outward expansion (pink shaded area and arrows in Figure 3a), which increases the surface area, is followed by the gradual emergence of the 50~100 nm nano-pyramids (Figure 3a, top row). In contrast, the {111} facets are resistant to oxidation and remained smooth (Figure 3a, middle row). As oxidation continues, however, etching begins to occur from the edges towards the center (blue shaded area and arrows in Figure 3a), effectively decreasing the surface area of the {111} facets over time. Etched areas can be seen at the junctions between the {111} facets (green shaded area in Figure 3a). Severe etching from prolonged oxidation causes the {100} facet to disappear and eventually form the flower-like structure (Figure 3a, bottom row). Because oxidation is initiated at different sites on the surface, small pyramids (pseudo-colored yellow in Figure 3a) form continuously on the smooth region all around the existing pyramids (pseudo-colored red in Figure 3a). This implies that it might be possible to create intriguing structures via intermittent (rather than continuous) oxidation.

Interestingly, when the oxidized diamond particles were used as seeds for diamond growth, the newly grown diamond particles reverted to their original, unsculpted shape (Figure S4). This proves that nanosculpting diamonds by air oxidation is the reversed process of diamond growth. This is a significant observation because being able to reverse the shape of diamonds without compromising their quality means that diamond structures could potentially be sculpted for specific applications and thereafter, recycled and reused by reverting to its original shape. Such shape reversible diamonds are also useful for exploring the fundamental properties and growth dynamics of diamond materials.

Our Monte Carlo simulations confirmed that different crystallographic facets etched at different rates. In our simulations, carbon atoms located at different corners, edges, and facets are categorized by the coordination number $n$, which is defined as the number of bonds the concerned carbon atom forms with neighboring carbon atoms (Figure 3b, top row). This number also represents the internal attractive energy to prohibit carbon atoms from oxidation. During the non-equilibrium oxidation process, etching of atoms on the {100} facet (where the coordination number of atoms is smaller than 3) is highly stochastic at the beginning. Over time, a significantly higher energy is needed to etch atoms with $n = 3$ that form. As more $n = 3$ atoms form on the surface, the small {111} facets eventually develop into pyramids on the initial {100} facets (Figure 3b, middle row). These pyramidal structures are stable and can enlarge and interconnect over time (Figure 3c). On the {111} facets, corner atoms with a lower coordination number are etched away faster than the inner atoms. Etched corner atoms lower the coordination number of adjacent atoms, causing them to also be etched (Figure 3b, bottom row). In this way, the atoms on the {111} facets are etched layer by layer from the edge to the center in a process known as step-recession [38]. The driving force for etching (represented by $n^*$), however, depends largely on oxidation conditions (temperature and oxygen concentration, see Methods for details). Higher level of oxidation conditions (*i.e.*, higher $n^*$) hastens etching (Figure 3d) irrespective of coordination numbers and this effect is even significant when external oxidation energy ($n^*$) is close to the inner attractive energy ($n$). For example, in the situation presented in Figure 3d, the external oxidation energy ($n^* = 1.95$~$2.00$) is significantly lower than the internal attractive energy of {111} atoms ($n = 3$) but comparable to the {100} atoms ($n = 2$). Therefore, the oxidation rate varies greatly for {100} atoms and has no significant impact on {111} atoms.

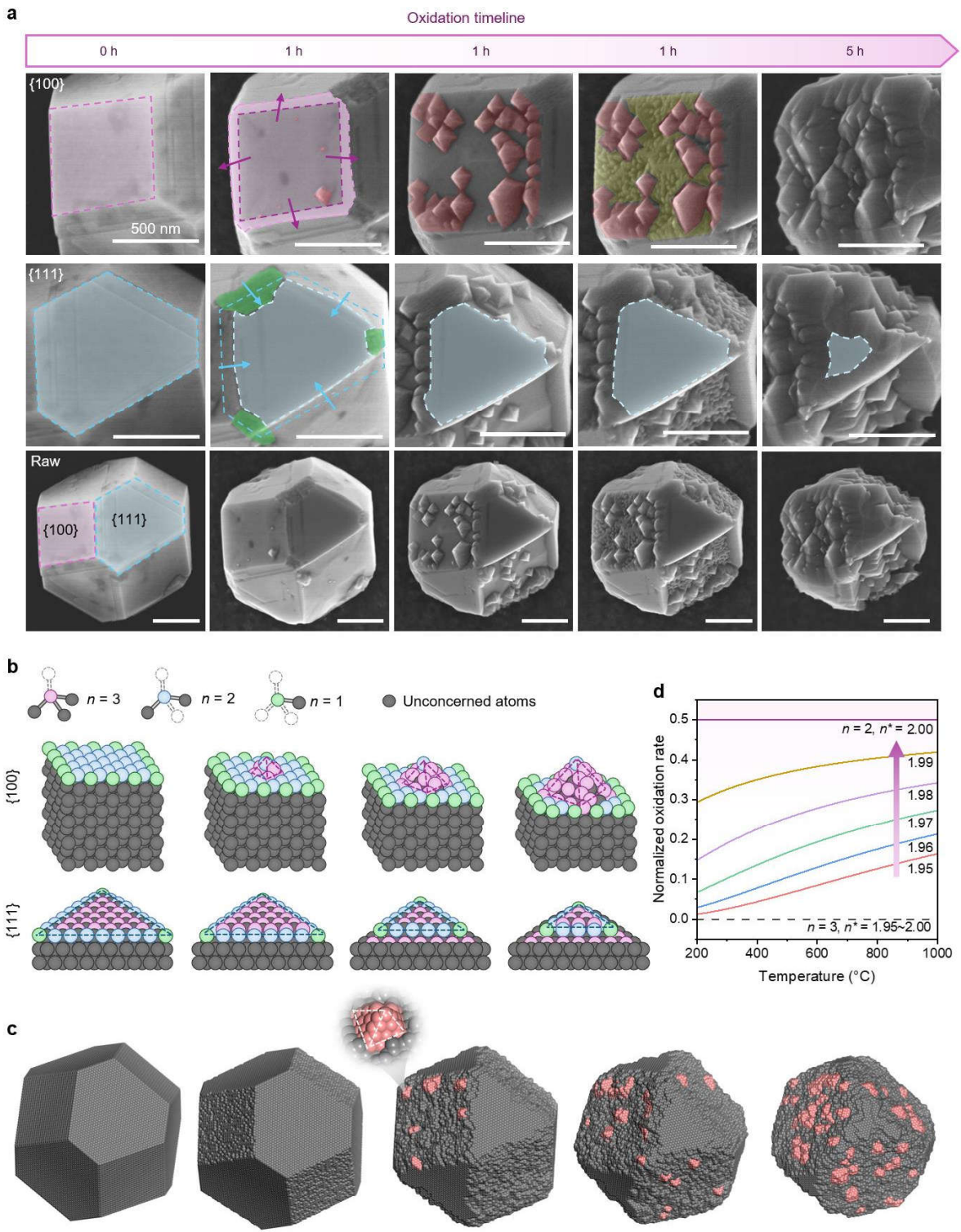

**Figure 3.** Different crystallographic facets in diamonds are etched at different rates during air oxidation. (a) SEM images showing the evolution of the {100} (purple dashed line) and {111}

(blue dashed line) facets on a regular shaped diamond microparticle. The {100} facet expands outward (purple arrows) to increase its surface area before pyramidal structures (red and yellow) form. Etching of the {111} facet occurs from the edges towards the center (blue arrows). (b) Schematic showing how pyramids form on the {100} surface, and how etching occurs on the {111} surface. (c) Simulated time-dependent evolution of a single diamond particle with a configuration similar to the one shown in (a). Inset: magnified view of the protruding pyramidal structures (red) on the diamond particle. (d) Simulated plots showing the oxidation rate of atoms with different coordination numbers ($n$ = 2 and 3) and external oxidation energy ($n^*$ = 1.95 to 2.00) at different temperatures (from 200 to 1000 °C). $n$ is the number of bonds the concerned carbon atom forms with neighboring atoms. $n^*$ represents external chemical potential.

**Hollow Shapes Form from Etching Crystal Defects Inside Out.**

While outward-to-inward etching on the facets explains how the protruding structures form on the particle surface, it cannot explain how the hollow particles shown in Figure 2e form. Upon careful re-examination of the original raw diamond particles shown in Figure 2c, it is apparent that these particles have irregular shapes, wrinkled lattice lines on the surface, and many crystal defects (Figure 4a). These features suggest that internal defects may also contribute to the shaping of the particle during air oxidation.

Cross-sectional dark field (DF) and high-resolution (HR) TEM shows the raw diamond is a polycrystal with numerous complex nanotwins and stacking faults (red rectangles in Figure 4a). Fast Fourier transform (FFT) pattern of the area with defects confirm stacking faults are present (inset in Figure 4a). Theoretically, the bonding energy of carbon atoms in defective regions is lower than in a perfect crystal.[39] This means that at high temperatures, carbon atoms at these defective sites are quickly and easily etched away to form a cavity. Indeed, cross-sectional HRTEM and FFT analysis of an oxidized (*i.e.,* hollow) diamond particle show no apparent crystal defects (Figure 4b). Oxidation had clearly etched away most of the defects inside the particle. Further macroscopic analysis using second harmonic generation (SHG) measurements confirmed the oxidation of internal defects in the diamond (Figure 4c). Only raw diamond particles displayed a strong SHG signal linked to the extent of internal defects in the diamond [40-41]. Together, these results show air oxidation etches away internal defects in diamond to form hollow particles.

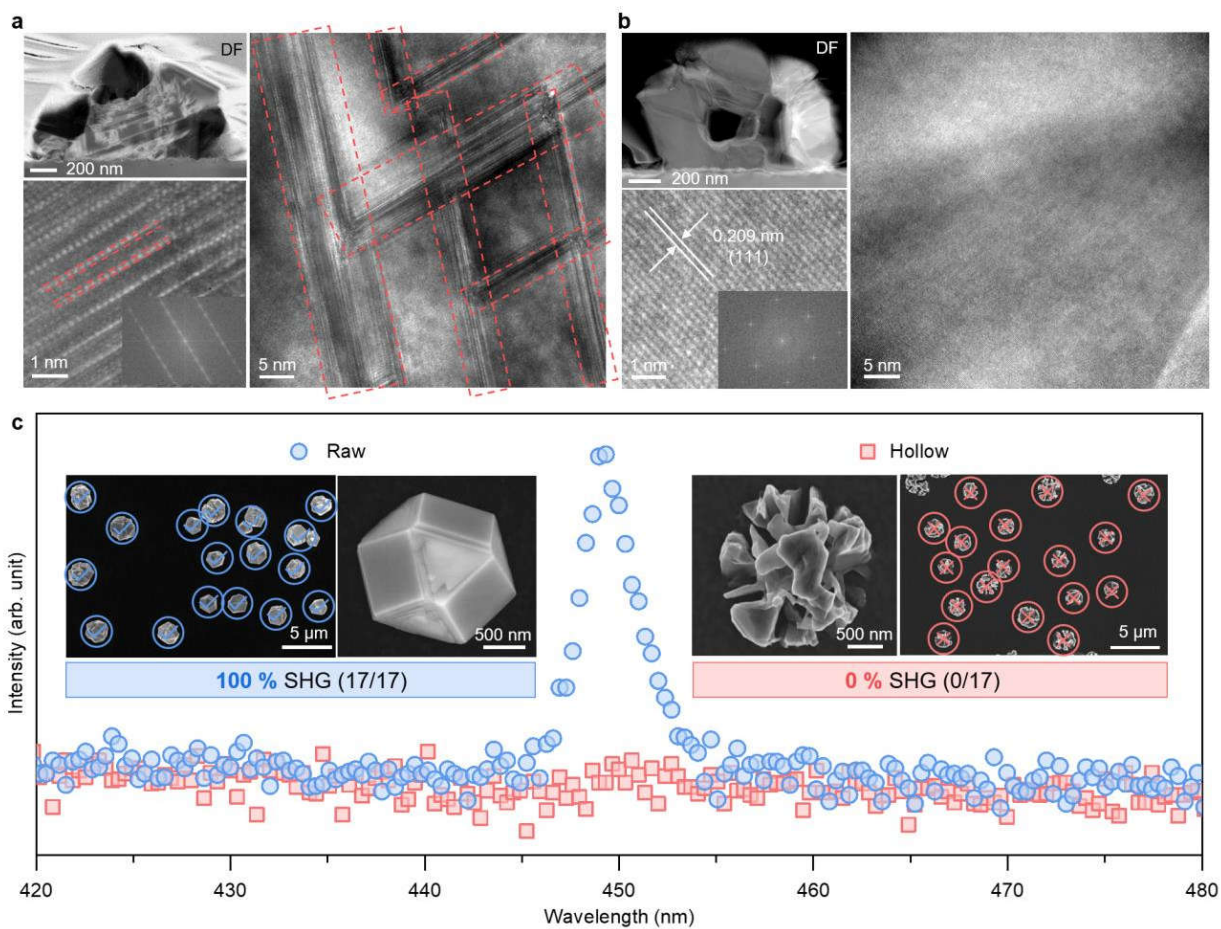

**Figure 4.** Air oxidation etches away crystal defects in diamond microparticles. Cross-sectional DF-TEM (top left), HR-TEM (right and bottom left) images and the corresponding FFT pattern

(bottom left inset) of (a) raw and (b) hollow diamond particles. The numerous nanotwins and stacking faults (red rectangles in (a)) in raw diamonds disappear after oxidation. Oxidized hollow diamond has a perfect diamond crystal structure. (c) SHG spectra of raw (blue circles) and hollow (red squares) diamond particles. Only raw diamonds display a strong SHG signal associated with crystal defects. Insets: SEM images (left: raw; right: hollow) of measured particles. In two random areas containing multiple diamond particles, all (17/17) raw diamonds have SHG signals while no signals (0/17) are detected in the hollow particles. The chance of obtaining SHG signals for raw and hollow particles is 100%:0%.

To further understand how the crystal defects are etched, we simulated the evolution of the polycrystalline diamond particle. Important factors such as the proportion of crystal facets, seed size, and crystal defects of the diamond were included in our model, so it accurately represents the real particle. For example, weak surfaces (with broken/missing atomic bonds) were introduced into the particle to represent its initial defects. Consistent with the experimental observations in Figure 2c, our model shows oxidation can eventually produce hollow and flower-like particles (Figure 5a, top row). Etching begins on the particle {100} surface and the weak interface (defects) inside the particle (Figure 5a, bottom row, magnified cross-sectional evolution views of two defective regions). Etching of internal defective atoms forms voids and holes inside the particle (red rectangle in Figure 5a). Removal of surface atoms leads to the removal of inner atoms along the direction of the crystal defects (blue rectangle in Figure 5a). Over time, the voids interconnect and coalesce, forming a flower-like particle with petals along different directions (stage iv, v in Figure 5a). Moreover, we have provided videos showing the dynamic oxidation process of this particle (Movie S2), along with 3D models showcasing the particle at different stages (Movies S3 to S5 for stages ii to iv). These visuals offer a more intuitive representation of the etching process of the particle and the corresponding 3D structures at each stage.

Meanwhile, the atomic ratios of the particle, such as the ratio of surface atoms to total atoms (Figure 5b) and the proportion of {111} ($n = 3$) atoms on the surface (Figure 5c), vary differently at different oxidation stages. With the initial removal of surface and internal defective atoms, these ratios show an increasing trend at the early stages. Particles with such increased ratios of surface atoms and {111} atoms are crucial for various novel applications[42-45] as they could display enhanced catalytic, optical, and/or electronic properties. It is worth noting that particles with different initial configurations exhibit distinct evolutions in these ratios during oxidation (see the blue and red lines in Figure 5b and 5c). At the same time, different oxidation conditions would also result in distinct trends in the atomic ratios, even for particles with the same initial configuration (Figure S5). These findings have important implications for the customization of diamond particles for different practical applications.

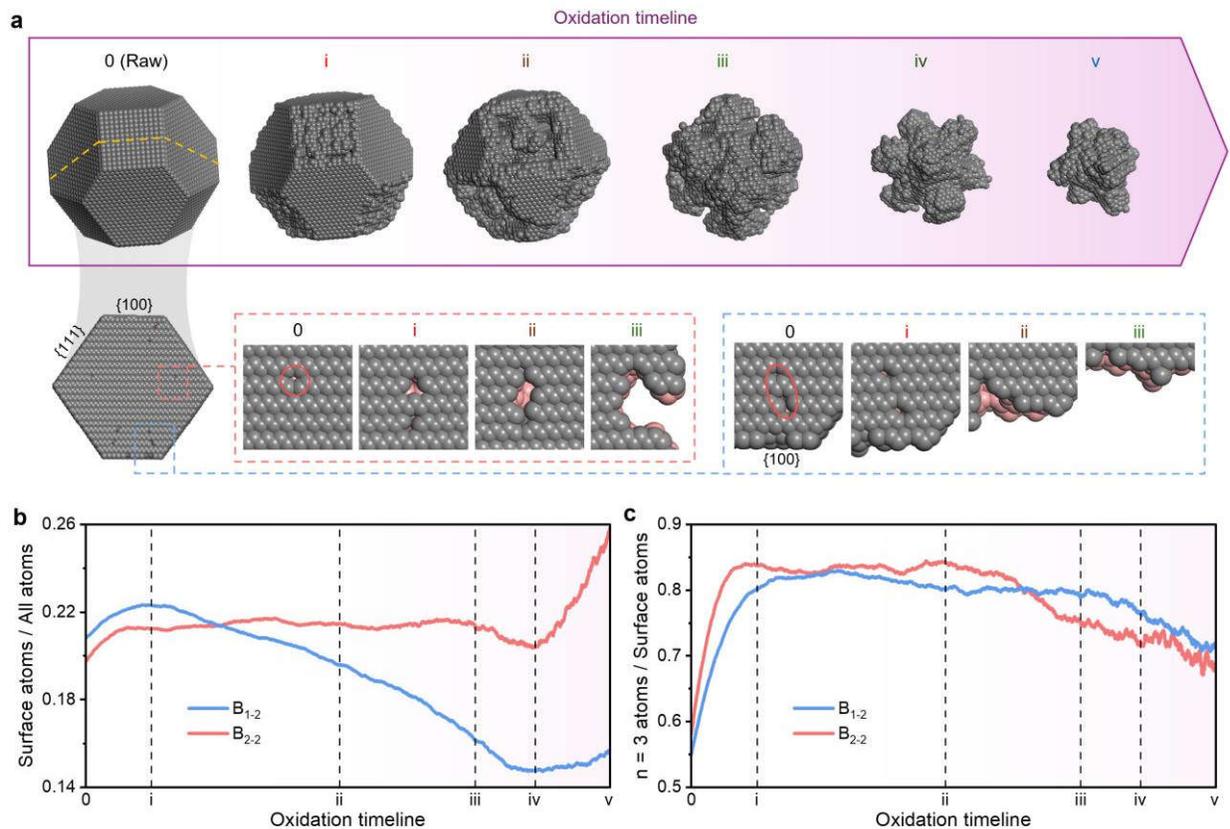

**Figure 5.** Inside out etching of crystallographic defects. (a) Simulated time-dependent evolution of a diamond model with crystal defects subjected to air oxidation, showing how these defects are etched from the inside out over time to form a hollow and flower-like particle (top row). Cross-sectional views of the particle (bottom row) in the original stage (0), showing the crystal defects (indicated by red circles) inside the particle, and the first three stages (i–iii) of the reaction, showing etching begins on crystal defects inside the particle (red rectangle), and removal of {100} surface atoms triggers the removal of inner defective atoms (blue rectangle). Continued oxidation produces significant voids in the crystal. The atoms located behind the cross-sectional layer are colored in red. Ratios of (b) the amount of surface atoms to the total amount of atoms and (c) the amount of $n = 3$ atoms to the surface atoms as a function of oxidation time for particles $B_{2-2}$ in (a) and $B_{1-2}$ in Figure 6.

**Simulated Library of Evolved Shapes Under Various Conditions**

Given the shape evolution of a diamond particle under air oxidation is largely determined by the particle's initial shape, seed size, proportion of crystal defects and facets, and the level of oxidation conditions, we constructed a shape library from our simulations to illustrate how these initial particle characteristics and oxidation conditions influence the trajectory of the particle shape (Figure 6). We found that particles with high ratio of {111} facets (*i.e.,* those with more $n = 3$ atoms) are more likely to be etched into spherical particles whereas those with more {100} facets tend to retain their ridged shapes (Figure 6, panel I). And crystals with a higher proportion of broken or missing atomic bonds (*i.e.*, crystal defects) can develop more etching paths into the particle (Figure 6, panel II). However, hollow structures form only if the diamond particle has more than 20% broken bonds and grows from small seeds. This suggests that it is harder to form extended etching paths in larger seeds (Figure 6, panel III). With the increase in oxidation conditions (*i.e.*, temperature and oxygen concentration), the probability of oxidation occurring along crystal defects rises, leading to less consolidated final shapes (Figure 6, panel IV). This indicates a clear transition in etching modes, shifting from outside-in to inside-out as the level of oxidation conditions increases. Additionally, 3D models demonstrating the shape transition of the same particle under different oxidation conditions have been provided in Movies S6 to S8 (the fourth stage of $B_{3-1}$ to $B_{3-3}$ in Figure 6).

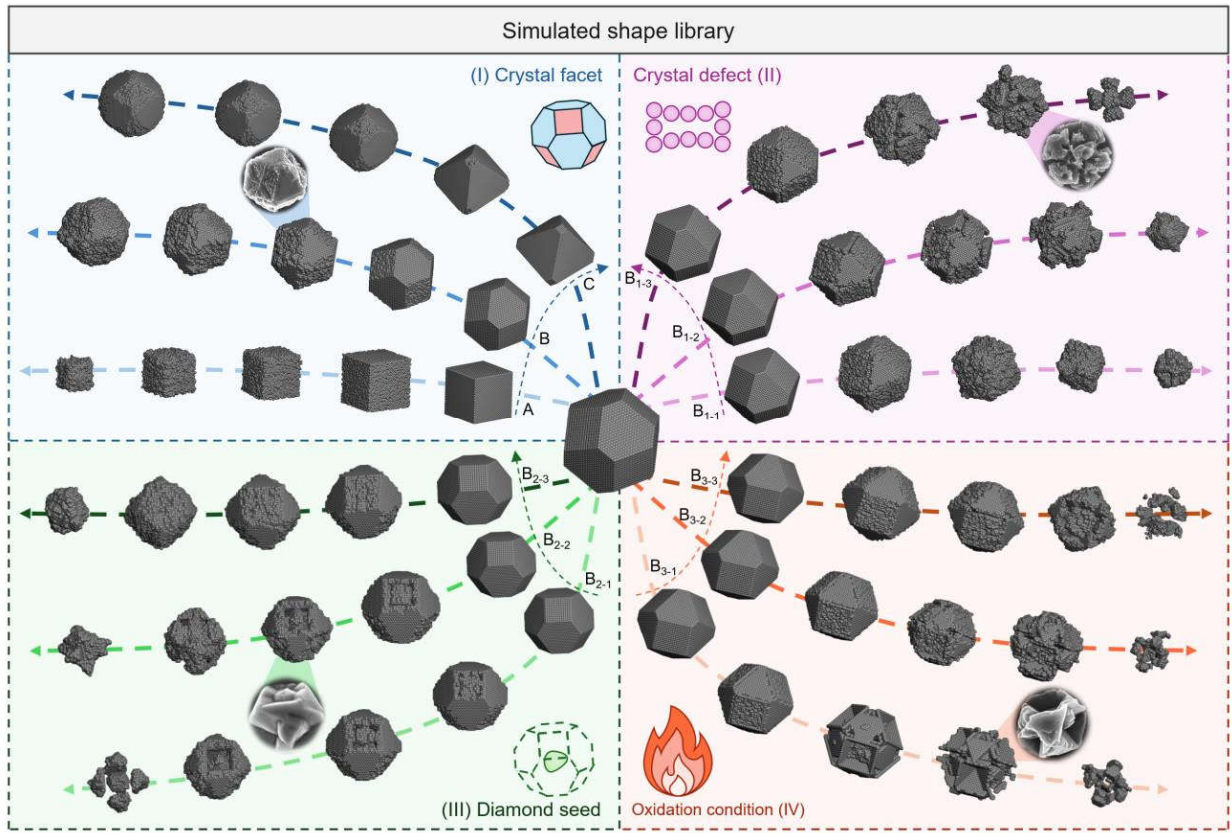

**Figure 6.** Shape library by Monte Carlo simulations guides diamond particle shape engineering. This library predicts the shapes that can be obtained based on the initial facet ratio (I, blue area), defect density (II, purple area) and seed size (III, green area) of a diamond particle, and the air oxidation conditions (IV, orange area) for the same diamond particle. I: The ratio of facets {111} to {100} increases from A to C (thin blue dashed arrow). II: Defect density (or proportion of weak bonds) with the same initial facet ratio (shape B) increases from $B_{1-1}$ to $B_{1-3}$ (thin purple dashed arrow). III: Seed size with the same initial facet ratio (shape B) increases from $B_{2-1}$ to $B_{2-3}$ (thin green dashed arrow). IV: The level of air oxidation condition for the same diamond particle (shape B) increases from $B_{3-1}$ to $B_{3-3}$ (thin orange dashed arrow). Thick dashed arrows track how the shape of each initial diamond particle changes upon air oxidation. Experimentally obtained particles are shown as insets. The same sculpting conditions were used for different initial particles (I, II and III), and the same particles were used for different oxidation conditions (IV).

Our simulated library is a guide for engineering diamond particles into desired shapes. Precisely engineering the shape of diamonds could potentially unlock a new era of diamond technology because the nanosculpted diamonds have new and/or more outstanding properties than the raw diamond particles. For example, the hollow diamond particles have ~ 11.38 times more surface features (as extracted by computer vision algorithm) and are less ordered than the original raw ones (Figure 7). Unlike our previous work,[36] where all particles (>2500) within a 200 μm × 200 μm area are required to produce a PUFs label, the abundant surface features and disordered characteristics of these nanosculpted hollow diamonds could potentially mean that only one particle is required to produce a PUFs label. Such single-particle PUFs are more secure and reliable than PUFs made from an ensemble of particles because it is harder to replicate or predict a single particle with unique and unpredictable response. Further, single-particle PUFs are easier to implement particularly in embedded systems and devices for the Internet of Things, where miniaturization is key.

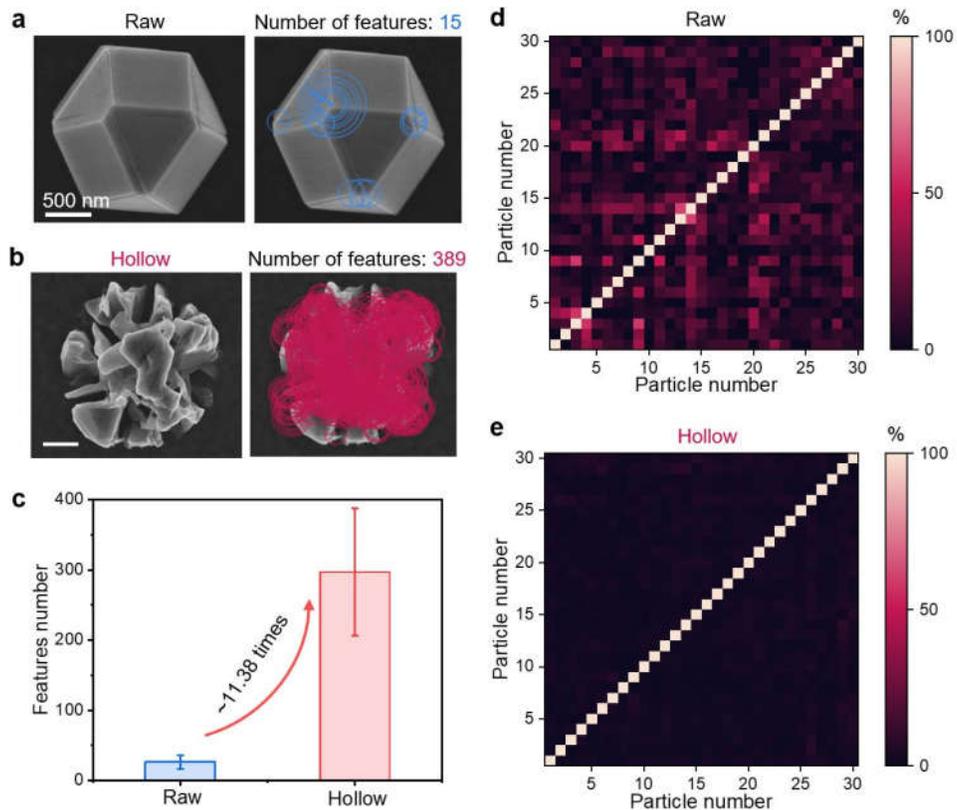

**Figure 7.** Single-particle level PUFs. Typical SEM images and the corresponding surface features extracted by computer vision algorithms from the SEM images for (a) raw and (b) hollow diamond particle. (c) Statistical analysis of the surface features extracted from 30 SEM images of both raw and hollow diamond particles, revealing that the hollow diamond particles exhibit approximately 11.38 times more surface features than the raw particles. Pairwise matches of each (d) raw and (e) hollow particle, with the color bar representing the similarity index. It is evident that the hollow particles demonstrate superior single-particle PUFs performance compared to the raw particles.

The altered features in our nanosculpted diamond particles also contribute to their unique optical properties, especially light scattering (Figure S6). The light scattering spectrum of the diamond particle changed as its shape evolved during air oxidation. Together with the different SHG responses of hollow and raw diamonds obtained in Figure 4c, these distinctive optical characteristics at the individual particle level indicates that the nanosculpted particles could potentially be used in high-end information encryption, high-density optical data storage, anti-counterfeiting measures, optical communications and other photonics applications requiring innovative light manipulation. Moreover, nanosculpted diamonds with higher surface ratios and exposed reactive {111} facets as shown in Figure 5b, 5c and S5 are advantageous for optics, biological, and catalytic applications.[42-45]

**CONCLUSIONS**

Our study shows simple air oxidation can be used to precisely shape diamond particles at scale. Polycrystalline diamond microparticles with diverse crystal facets and internal defects that display different reactivities are sculpted into complex shapes such as sphere, twisted surface, pyramidal islands, inverted pyramid, flower, and hollow polygon. We show that protruding structures are a result of etching crystal facets from the outside in while hollow structures are formed from etching crystal defects from the inside out. Compared to the original diamond particles, the sculpted diamonds have more and finer features that could potentially improve existing applications or be applied in new ways. Unlike existing shape engineering techniques, air oxidation is simple, economical, and scalable. Multiple diamond particles can be sculpted simultaneously using standard equipment and mild experimental conditions. Using Monte Carlo simulations, we established a shape library to guide anyone interested in producing large quantities of diamond particles with well-defined shapes. The library predicts the shape based on the initial characteristics of the diamond particle, ensuring the right oxidation conditions are used to obtain each desired shape. We also discovered that air oxidizing diamonds is a reversed process of diamond growth. And this means that diamonds could, in the future, be sculpted for specific applications and recycled or reused thereafter. Having a sustainable form of high-quality bespoke diamonds opens new possibilities across various fields including photonics, electronics, catalysis, quantum and information technologies.

**EXPERIMENTAL SECTION**

**The Fabrication of Diamond Microparticles.** Following our previous work,[35-36] 50 nm salt-assisted air-oxidized nanodiamonds (SAAO NDs)[46] were used as CVD seeds for the growth of diamond microparticles. Specifically, the process involved air oxidizing a mixture comprising 10 mg of NDs (50 nm, HPHT, PolyQolor) and 50 mg of sodium chloride (NaCl, 99.5%, Sigma-Aldrich) at 500 °C for 1 hour. Following oxidation, the NaCl residue was removed by washing with deionized (DI) water through centrifugation. Then, 0.1 mg/mL SAAO NDs in DI water was prepared for the subsequent diamond growth. The SAAO NDs suspension was spin-coated onto the Si substrate, and diamond growth was carried out for 80 minutes in a microwave-plasma assisted chemical vapor deposition (MPCVD) system (Seki 6350) under constant power (3400 W), pressure (85 torr), and temperature (920 °C) using a gas mixture of $H_2/CH_4$ (94/6).

**Air Oxidation of Diamond Microparticles.** The CVD-grown diamond microparticles were subjected to oxidation in air to manipulate their shapes. By varying the oxidation temperature range from 500 to 700 °C and duration from 1 to 20 hours, a diverse array of diamond particle shapes was successfully achieved.

**3D Shape Reconstruction from SEM Images.** The 3D shape model of the hollow diamond particle was constructed by reconstructing SEM images of the particle captured at various tilting angles (from 0° to 42° with 6° steps). Each titling angle set of SEM images was obtained with a 60° rotational step covering a full 360° rotation. The subsequent 3D shape reconstruction process was carried out using Meshroom (Https://github.Com/alicevision/meshroom), a freely available open-source software that utilizes the AliceVision Photogrammetric Computer Vision framework.

**Study On the Shape Evolution of Single Diamond Microparticle.** One selected diamond microparticle exhibiting a regular shape with clear {111} and {100} crystal facets was observed and located under SEM. Subsequently, the Si substrate carrying this specific particle underwent air oxidation within a furnace at 650 °C for a duration of 1 hour. This process, involving oxidation at 650 °C for 1 hour followed by SEM analysis, was repeated twice. Finally, the sample underwent a prolonged air oxidation period of 5 hours at 650 °C, after which it was once again viewed under SEM.

**Regrowth Diamond from The Reshaped Diamond Particles.** The diamond particles were first reshaped into flower-like shapes through a serials of air oxidation steps. Then, the Si substrate containing the reshaped diamond particles was directly put into the CVD chamber for diamond growth, which lasted for a period of 40 minutes, under the same conditions of the initial diamond growth.

**Theoretical Simulations.** Kinetic Monte Carlo model: The state distribution of the diamond atoms is presented by the probability function:

$$P(C) \propto e^{-\beta H} \tag{1}$$

where $\beta = 1/(k_\text{B}T)$ with $k_\text{B}$ being the Boltzmann constant and $T$ being the temperature. The equilibrium fluctuations in a grand canonical ensemble are determined by the distribution $C = \{\rho_1, \rho_2, \cdots\}$, where $\rho_i$ represents the state of the individual atom $i$ in presence ($\rho_i = 1$) or absence ($\rho_i = 0$). $H$ is the Hamiltonian, which is expressed as:

$$H = -\mu \sum_i \rho_i - \epsilon \sum_{\langle i,j \rangle} \rho_i \rho_j \qquad (2)$$

in which $\mu$ is the externally imposed chemical potential energy and $\epsilon$ is the attractive energy between each nearest couple of atoms. In the situation of diamond oxidation, $\mu$ represents the potential energy to form C-O bond and $\epsilon$ represents the potential energy of C-C bond. Here a relationship can be established to measure the relative value between external potential energy and internal bond energy as:

$$\mu = -n^* \epsilon \qquad (3)$$

where $n^*$ represents the standard external chemical potential that quantitively measures the relative amount of two potential energies. For general lattice, it exhibits coexistence between macroscopic high-density and low-density phases when $n^* = Z/2$, in which $Z$ is the maximum number of nearest neighbors that a lattice site can have. For diamond, $Z = 4$. The external chemical potential greatly depends on the oxidation conditions (temperature and oxygen concentration). For example, increasing the temperature and oxygen concentration can significantly increase the external chemical potential by altering the absorption structure on the surface and crystal defects.[47-49]

In the above scheme of distribution probabilities, when the state of an atom $i$ changes from $\rho_i$ to $1 - \rho_i$ (and the corresponding system state changes from $C$ to $C'$), the probability can be determined as:

$$P(C \to C') = 1/\{1 + e^{\beta[H(C')-H(C)]}\} \quad (4)$$

Combining the Equations (1) to (4), when a diamond atom $i$ is oxidated from the particle ($\rho_i = 1 \to \rho_i = 0$), the probability is then derived as:

$$P(C \to C') = 1/[1 + e^{\beta\epsilon(n-n^*)}] \quad (5)$$

and thus, the reaction rate can be expressed as:

$$k = w_0 P(C \to C') = w_0/[1 + e^{\beta\epsilon(n-n^*)}] \quad (6)$$

in which $w_0$ is a constant quantifying the reaction rate and coordination number $n$ counts the total amount of the nearest atoms to the concerned atom. In Figure 3$E$ the etching rate was normalized by deriving $w_0 = 1$.

**Simulation Procedure.** Step 1. Generate initial configuration of the diamond particles, where the surfaces were cut from tightly arranged diamond grain cells. Different crystal facets were generated by changing cutting directions. Step 2. Weak interfaces were generated outside a central sphere region (diamond seed) along the direction from the particle center to the boundaries of different surfaces. Coordination number of atoms on the weak interfaces was lowered to simulate the crystal defects inside the particles. Defect density was determined by changing the proportion of atoms with lowered coordination number. Seed size was determined by changing the radius of the central sphere region. Oxidation condition level was determined by changing the external chemical potential, where a higher value of $n^*$ represented a higher oxidation level. Step 3. Compute the probability of each atom and determine one atom to be etched in each step based on the probability distribution. Length of timestep was determined by the etching rate in Equation (6). Step 4. Remove the etched atom and update the neighboring information of the remaining carbon atoms. Step 5. Repeat steps 3 to 4. The end condition was set as the etched atom number or oxidation time. In this work, the simulated atom numbers within one diamond particle varied from $5 \times 10^4$ to $4 \times 10^5$.

**Computer Vision-Based Feature Extraction and Matching of The Diamond Particles.** The feature extraction and matching algorithm has been implemented by utilizing computer vision algorithms from OpenCV python library. The scale-invariant feature transform (SIFT) based algorithm (ORB was also tested and yielding similar results) was employed for extracting and matching local surface features of raw and hollow shaped diamond particles. The matching score (or similarity) was calculated based on successful matching of feature points. A set of 30 SEM images for each kind of particle, including both raw and hollow, was utilized for analysis.

**Characterizations.** Scanning electron microscope (SEM, Hitachi S4800) and scanning transmission electron microscope (STEM, Thermo Scientific Talos F200X) were employed to analyze the morphology features of the diamond samples. High-resolution TEM (HRTEM) allows the observation of the arrangement of diamond carbon atoms at nanoscale. Focus ion beam (FIB, FEI Quanta 200 3D Dual Beam) was used to prepare cross-sectional TEM samples. The crystallinity of the grown diamond was analyzed using X-ray diffractometer (XRD, D8 Advance, Bruker). The Raman spectrum of the sample was measured by a Renishaw InVia Raman microscope. The SHG measurements were carried out by a home-built optical system. A pulsed laser of 900 nm wavelength was generated using a femtosecond laser source (Chameleon) and the SHG signal of 450 nm was recorded by a spectrograph (Andor, SR500i). Single-particle dark-field scattering spectrum was recorded on an Olympus upright microscope that consists of a quartz-tungsten-halogen lamp (100W), a spectrograph (Acton, SpectraPro 2360i), and a charge-coupled device camera (Princeton Instruments, Pixis 400).

**ASSOCIATED CONTENT**

**Supporting Information**.

SEM images of the diamond particles after different nanosculpting conditions; SEM images showing the 100% yield and uniformity of the nanosculpting process; SEM images of the regrowth-diamond from the reshaped diamond particles; effect of oxidation conditions on atomic ratios of the particle; the evolution of scattering spectrum of single particle upon air oxidation nanosculpting (PDF)

Movie showing the 3D reconstructed model of hollow particle shown in Figure 2e; movies showing dynamic oxidation process and 3D models of stages ii – iv of the particle shown in Figure 5a; movies showing the 3D models of stage iv of the particles $B_{3-1}$ – $B_{3-3}$ shown in Figure 6 (MP4)

**AUTHOR INFORMATION**

**Corresponding Author**

* To whom correspondence should be addressed:

Y.L. (ylin@hku.hk) and Z.Q.C. (zqchu@eee.hku.hk).

**Author Contributions**

‡T.T.Z. and F.Q.S. contributed equally to this work. Z.Q.C. and T.T.Z. conceived the ideas and designed the experiments. Z.Q.C. and Y.L. supervised the project. T.T.Z., F.Q.S. Z.Q.C., and Y.L., wrote the manuscript. T.T.Z. conducted the experiments and collected the data. F.Q.S. carried out the Monte Carlo simulations under the supervision of Y.L.. Z.Q.W. and Q.W. supported the MPCVD system. Y.C.L. and N.W. supported the 3D shape reconstruction. Y.R.W. and D.Y.L. supported the SHG measurements. J.W. and L.S. supported the scattering spectrum measurements. All authors discussed the results and contributed to the manuscript.

**Notes**

Z.Q.C, T.T.Z. Q.W., Z.Q.W. and K.W.L. are inventors on the Chinese Invention Patent Application No. 202310062039.5 filed on 18 January 2023 and US Non-Provisional Patent Application No. 18/416,325 filed on 18 January 2024. The remaining authors declare no other competing interests.

**ACKNOWLEDGMENT**

Z.Q.C. acknowledges the financial support from the National Natural Science Foundation of China (NSFC) and the Research Grants Council (RGC) of Hong Kong Joint Research Scheme (Project No. N_HKU750/23); RGC Theme-based Research Scheme (Project No. T45-701/22-R); HKU Seed Fund; and the Health@InnoHK program of the Innovation and Technology Commission of the Hong Kong SAR Government. Y.L. thanks the financial support from the RGC General Research Fund (GRF, Project No. 17210520), the Health@InnoHK program of the Innovation and Technology Commission of the Hong Kong SAR Government, and the National Natural Science Foundation of China (Project No. 12272332). Q.W. acknowledges the financial support of Guangdong Major Project of Basic and Applied Basic Research 2023B0303000012. D.Y.L. acknowledges the financial support of the Research Grants Council of Hong Kong through a Collaborative Research Equipment Grant (C1015-21EF) and an Area of Excellence grant (AoE/P-701/20). The authors are grateful to Prof. Jianfang Wang and Dr. Jiapeng Zheng (Department of Physics, The Chinese University of Hong Kong) for help with the single-particle dark-field scattering spectrum measurement in the current work.

**REFERENCES**

1. Mochalin, V. N.; Shenderova, O.; Ho, D.; Gogotsi, Y., The properties and applications of nanodiamonds. *Nat. Nanotechnol.* **2011,** *7* (1), 11−23.

2. Huang, Q.; Yu, D.; Xu, B.; Hu, W.; Ma, Y.; Wang, Y.; Zhao, Z.; Wen, B.; He, J.; Liu, Z.; Tian, Y., Nanotwinned diamond with unprecedented hardness and stability. *Nature* **2014,** *510* (7504), 250−253.

3. Miller, B. S.; Bezinge, L.; Gliddon, H. D.; Huang, D.; Dold, G.; Gray, E. R.; Heaney, J.; Dobson, P. J.; Nastouli, E.; Morton, J. J. L.; McKendry, R. A., Spin-enhanced nanodiamond biosensing for ultrasensitive diagnostics. *Nature* **2020,** *587* (7835), 588−593.

4. Shilkin, D. A.; Shcherbakov, M. R.; Lyubin, E. V.; Katamadze, K. G.; Kudryavtsev, O. S.; Sedov, V. S.; Vlasov, I. I.; Fedyanin, A. A., Optical magnetism and fundamental modes of nanodiamonds. *ACS Photonics* **2017,** *4* (5), 1153−1158.

5. Reineck, P.; Trindade, L. F.; Havlik, J.; Stursa, J.; Heffernan, A.; Elbourne, A.; Orth, A.; Capelli, M.; Cigler, P.; Simpson, D. A.; Gibson, B. C., Not all fluorescent nanodiamonds are created equal: A comparative study. *Part. Part. Syst. Charact.* **2019,** *36* (36), 1900009.

6. Lin, Y.; Sun, X.; Su, D. S.; Centi, G.; Perathoner, S., Catalysis by hybrid $sp^2/sp^3$ nanodiamonds and their role in the design of advanced nanocarbon materials. *Chem. Soc. Rev.* **2018,** *47* (22), 8438−8473.

7. Chu, Z.; Zhang, S.; Zhang, B.; Zhang, C.; Fang, C. Y.; Rehor, I.; Cigler, P.; Chang, H. C.; Lin, G.; Liu, R.; Li, Q., Unambiguous observation of shape effects on cellular fate of nanoparticles. *Sci. Rep.* **2014,** *4*, 4495.

8. Im, S. W.; Zhang, D.; Han, J. H.; Kim, R. M.; Choi, C.; Kim, Y. M.; Nam, K. T., Investigating chiral morphogenesis of gold using generative cellular automata. *Nat. Mater.* **2024,** *23* (7), 977−983.

9. Chen, Q.; Bae, S. C.; Granick, S., Directed self-assembly of a colloidal kagome lattice. *Nature* **2011,** *469* (7330), 381−384.

10. Zhou, S.; Li, J.; Lu, J.; Liu, H.; Kim, J. Y.; Kim, A.; Yao, L.; Liu, C.; Qian, C.; Hood, Z. D.; Lin, X.; Chen, W.; Gage, T. E.; Arslan, I.; Travesset, A.; Sun, K.; Kotov, N. A.; Chen, Q., Chiral assemblies of pinwheel superlattices on substrates. *Nature* **2022,** *612* (7939), 259−265.

11. Kumar, P.; Vo, T.; Cha, M.; Visheratina, A.; Kim, J. Y.; Xu, W.; Schwartz, J.; Simon, A.; Katz, D.; Nicu, V. P.; Marino, E.; Choi, W. J.; Veksler, M.; Chen, S.; Murray, C.; Hovden, R.; Glotzer, S.; Kotov, N. A., Photonically active bowtie nanoassemblies with chirality continuum. *Nature* **2023,** *615* (7952), 418−424.

12. Kronenfeld, J. M.; Rother, L.; Saccone, M. A.; Dulay, M. T.; DeSimone, J. M., Roll-to-roll, high-resolution 3D printing of shape-specific particles. *Nature* **2024,** *627* (8003), 306−312.

13. Zhou, W.; Lim, Y.; Lin, H.; Lee, S.; Li, Y.; Huang, Z.; Du, J. S.; Lee, B.; Wang, S.; Sanchez-Iglesias, A.; Grzelczak, M.; Liz-Marzan, L. M.; Glotzer, S. C.; Mirkin, C. A., Colloidal quasicrystals engineered with DNA. *Nat. Mater.* **2024,** *23* (3), 424−428.

14. Wintersinger, C. M.; Minev, D.; Ershova, A.; Sasaki, H. M.; Gowri, G.; Berengut, J. F.; Corea-Dilbert, F. E.; Yin, P.; Shih, W. M., Multi-micron crisscross structures grown from DNA-origami slats. *Nat. Nanotechnol.* **2023,** *18* (3), 281−289.

15. Posnjak, G.; Yin, X.; Butler, P.; Bienek, O.; Dass, M.; Lee, S.; Sharp, I. D.; Liedl, T., Diamond-lattice photonic crystals assembled from DNA origami. *Science* **2024,** *384* (6697), 781−785.

16. Liu, H.; Matthies, M.; Russo, J.; Rovigatti, L.; Narayanan, R. P.; Diep, T.; McKeen, D.; Gang, O.; Stephanopoulos, N.; Sciortino, F.; Yan, H.; Romano, F.; Sulc, P., Inverse design of a pyrochlore lattice of DNA origami through model-driven experiments. *Science* **2024,** *384* (6697), 776−781.

17. Wang, Q.; Wang, Z.; Li, Z.; Xiao, J.; Shan, H.; Fang, Z.; Qi, L., Controlled growth and shape-directed self-assembly of gold nanoarrows. *Sci. Adv.* **2017,** *3* (10), e1701183.

18. Huang, L.; Liu, M.; Lin, H.; Xu, Y.; Wu, J.; Dravid, V. P.; Wolverton, C.; Mirkin, C. A., Shape regulation of righ-index facet nanoparticles by dealloying. *Science* **2019,** *365* (6458), 1159−1163.

19. Xia, Y.; Nelli, D.; Ferrando, R.; Yuan, J.; Li, Z. Y., Shape control of size-selected naked platinum nanocrystals. *Nat. Commun.* **2021,** *12* (1), 3019.

20. Lyu, D.; Xu, W.; Payong, J. E. L.; Zhang, T.; Wang, Y., Low-dimensional assemblies of metal-organic framework particles and mutually coordinated anisotropy. *Nat. Commun.* **2022,** *13* (1), 3980.

21. Kim, M.; Xin, R.; Earnshaw, J.; Tang, J.; Hill, J. P.; Ashok, A.; Nanjundan, A. K.; Kim, J.; Young, C.; Sugahara, Y.; Na, J.; Yamauchi, Y., Mof-derived nanoporous carbons with diverse tunable nanoarchitectures. *Nat. Protoc.* **2022,** *17* (12), 2990−3027.

22. Shen, K.; Zhang, L.; Chen, X.; Liu, L.; Zhang, D.; Han, Y.; Chen, J.; Long, J.; Luque, R.; Li, Y.; Chen, B., Ordered macro-microporous metal-organic framework single crystals. *Science* **2018,** *359* (6372), 206−210.

23. Aharonovich, I.; Greentree, A. D.; Prawer, S., Diamond photonics. *Nat. Photonics* **2011,** *5* (7), 397−405.

24. Momenzadeh, S. A.; Stohr, R. J.; de Oliveira, F. F.; Brunner, A.; Denisenko, A.; Yang, S.; Reinhard, F.; Wrachtrup, J., Nanoengineered diamond waveguide as a robust bright platform for nanomagnetometry using shallow nitrogen vacancy centers. *Nano Lett.* **2015,** *15* (1), 165−169.

25. Xie, L.; Zhou, T. X.; Stohr, R. J.; Yacoby, A., Crystallographic orientation dependent reactive ion etching in single crystal diamond. *Adv. Mater.* **2018,** *30* (11), 1705501.

26. Atikian, H. A.; Sinclair, N.; Latawiec, P.; Xiong, X.; Meesala, S.; Gauthier, S.; Wintz, D.; Randi, J.; Bernot, D.; DeFrances, S.; Thomas, J.; Roman, M.; Durrant, S.; Capasso, F.; Loncar, M., Diamond mirrors for high-power continuous-wave lasers. *Nat. Commun.* **2022,** *13* (1), 2610.

27. Gong, Y.; Luo, D.; Choe, M.; Kim, Y.; Ram, B.; Zafari, M.; Seong, W. K.; Bakharev, P.; Wang, M.; Park, I. K.; Lee, S.; Shin, T. J.; Lee, Z.; Lee, G.; Ruoff, R. S., Growth of diamond in liquid metal at 1 atm pressure. *Nature* **2024,** *629* (8011), 348−354.

28. Taylor, A. C.; Edgington, R.; Jackman, R. B., Patterning of nanodiamond tracks and nanocrystalline diamond films using a micropipette for additive direct-write processing. *ACS Appl. Mater. Interfaces* **2015,** *7* (12), 6490−6495.

29. Osswald, S.; Yushin, G.; Mochalin, V.; Kucheyev, S. O.; Gogotsi, Y., Control of $sp^2/sp^3$ carbon ratio and surface chemistry of nanodiamond powders by selective oxidation in air. *J. Am. Chem. Soc.* **2006,** *128* (35), 11635−11642.

30. Stehlik, S.; Varga, M.; Ledinsky, M.; Jirasek, V.; Artemenko, A.; Kozak, H.; Ondic, L.; Skakalova, V.; Argentero, G.; Pennycook, T.; Meyer, J. C.; Fejfar, A.; Kromka, A.; Rezek, B., Size and purity control of HPHT nanodiamonds down to 1 nm. *J. Phys. Chem. C* **2015,** *119* (49), 27708−27720.

31. Tisler, J.; Balasubramanian, G.; Naydenov, B.; Kolesov, R.; Grotz, B.; Reuter, R.; Boudou, J.-P.; Curmi, P. A.; Sennour, M.; Thorel, A., Fluorescence and spin properties of defects in single digit nanodiamonds. *ACS Nano* **2009,** *3* (7), 1959−1965.

32. Gracio, J. J.; Fan, Q. H.; Madaleno, J. C., Diamond growth by chemical vapour deposition. *J. Phys. D: Appl. Phys.* **2010,** *43* (37), 374017.

33. Nemeth, P.; McColl, K.; Garvie, L. A. J.; Salzmann, C. G.; Murri, M.; McMillan, P. F., Complex nanostructures in diamond. *Nat. Mater.* **2020,** *19* (11), 1126−1131.

34. Zheng, Q.; Shi, X.; Jiang, J.; Mao, H.; Montes, N.; Kateris, N.; Reimer, J. A.; Wang, H.; Zheng, H., Unveiling the complexity of nanodiamond structures. *Proc. Natl. Acad. Sci. U. S. A.* **2023,** *120* (23), 2301981120.

35. Zhang, T.; Gupta, M.; Jing, J.; Wang, Z.; Guo, X.; Zhu, Y.; Yiu, Y. C.; Hui, T. K. C.; Wang, Q.; Li, K. H.; Chu, Z., High-quality diamond microparticles containing SiV centers grown by chemical vapor deposition with preselected seeds. *J. Mater. Chem. C* **2022,** *10* (37), 13734–13740.

36. Zhang, T.; Wang, L.; Wang, J.; Wang, Z.; Gupta, M.; Guo, X.; Zhu, Y.; Yiu, Y. C.; Hui, T. K. C.; Zhou, Y.; Li, C.; Lei, D.; Li, K. H.; Wang, X.; Wang, Q.; Shao, L.; Chu, Z., Multimodal dynamic and unclonable anti-counterfeiting using robust diamond microparticles on heterogeneous substrate. *Nat. Commun.* **2023,** *14* (1), 2507.

37. Arppe, R.; Sørensen, T. J., Physical unclonable functions generated through chemical methods for anti-counterfeiting. *Nat. Rev. Chem.* **2017,** *1* (4), 0031.

38. Ye, X.; Jones, M. R.; Frechette, L. B.; Chen, Q.; Powers, A. S.; Ercius, P.; Dunn, G.; Rotskoff, G. M.; Nguyen, S. C.; Adiga, V. P.; Zettl, A.; Rabani, E.; Geissler, P. L.; Alivisatos, A. P., Single-particle mapping of nonequilibrium nanocrystal transformations. *Science* **2016,** *354* (6314), 874−877.

39. Shenderova, O.; Brenner, D.; Omeltchenko, A.; Su, X.; Yang, L., Atomistic modeling of the fracture of polycrystalline diamond. *Phys. Rev. B* **2000,** *61* (6), 3877.

40. Abulikemu, A.; Kainuma, Y.; An, T.; Hase, M., Second-harmonic generation in bulk diamond based on inversion symmetry breaking by color centers. *ACS Photonics* **2021,** *8* (4), 988−993.

41. Abulikemu, A.; Kainuma, Y.; An, T.; Hase, M., Temperature-dependent second-harmonic generation from color centers in diamond. *Opt. Lett.* **2022,** *47* (7), 1693−1696.

42. Regan, B.; Kim, S.; Ly, A. T. H.; Trycz, A.; Bray, K.; Ganesan, K.; Toth, M.; Aharonovich, I., Photonic devices fabricated from (111)‐oriented single crystal diamond. *InfoMat* **2020,** *2* (6), 1241−1246.

43. Damle, V.; Wu, K.; De Luca, O.; Ortí-Casañ, N.; Norouzi, N.; Morita, A.; de Vries, J.; Kaper, H.; Zuhorn, I. S.; Eisel, U.; Vanpoucke, D. E. P.; Rudolf, P.; Schirhagl, R., Influence of diamond crystal orientation on the interaction with biological matter. *Carbon* **2020,** *162*, 1−12.

44. Wang, H.; Tzeng, Y. K.; Ji, Y.; Li, Y.; Li, J.; Zheng, X.; Yang, A.; Liu, Y.; Gong, Y.; Cai, L.; Li, Y.; Zhang, X.; Chen, W.; Liu, B.; Lu, H.; Melosh, N. A.; Shen, Z. X.; Chan, K.; Tan, T.; Chu, S.; Cui, Y., Synergistic enhancement of electrocatalytic $CO_2$ reduction to $C_2$ oxygenates at nitrogen-doped nanodiamonds/cu interface. *Nat. Nanotechnol.* **2020,** *15* (2), 131−137.

45. Sobaszek, M.; Brzhezinskaya, M.; Olejnik, A.; Mortet, V.; Alam, M.; Sawczak, M.; Ficek, M.; Gazda, M.; Weiss, Z.; Bogdanowicz, R., Highly occupied surface states at deuterium-grown boron-doped diamond interfaces for efficient photoelectrochemistry. *Small* **2023**, 2208265.

46. Zhang, T.; Ma, L.; Wang, L.; Xu, F.; Wei, Q.; Wang, W.; Lin, Y.; Chu, Z., Scalable fabrication of clean nanodiamonds *via* salt-assisted air oxidation: Implications for sensing and imaging. *ACS Appl. Nano Mater.* **2021,** *4* (9), 9223−9230.

47. Pu, J.-C.; Wang, S.-F.; Sung, J. C., High-temperature oxidation behavior of nanocrystalline diamond films. *J. Alloys Compd.* **2010,** *489* (2), 638−644.

48. Gamo, H.; Gamo, M. N.; Nakagawa, K.; Ando, T., Surface potential change by oxidation of the chemical vapor deposited diamond (001) surface. *J. Phys.: Conf. Ser.* **2007,** *61*, 327−331.

49. Pehrsson, P. E.; Mercer, T. W., Oxidation of heated diamond C(100):H surfaces. *Surf. Sci.* **2000,** *460* (1-3), 74−90.

**Graphical Table of Contents**

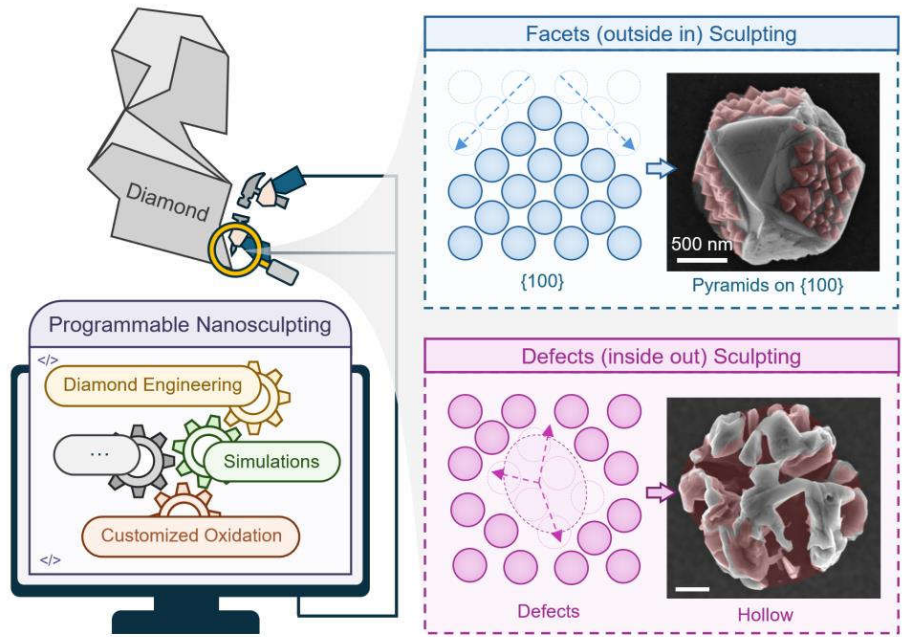

Supporting Information

# Scalable Reshaping of Diamond Particles via Programmable Nanosculpting

Tongtong Zhang,[1, 2, ‡] Fuqiang Sun,[3, ‡] Yaorong Wang,[4] Yingchi Li,[1] Jing Wang,[5] Zhongqiang Wang,[6] Kwai Hei Li,[7] Ye Zhu,[8] Qi Wang,[6] Lei Shao,[5] Ngai Wong,[1] Dangyuan Lei,[4] Yuan Lin,[2, 3, *] and Zhiqin Chu[1, 2, 9, *]

[1] Department of Electrical and Electronic Engineering, The University of Hong Kong, Pokfulam Road, Hong Kong, China

[2] Advanced Biomedical Instrumentation Centre, Hong Kong Science Park, Shatin, New Territories, Hong Kong, China

[3] Department of Mechanical Engineering, The University of Hong Kong, Pokfulam Road, Hong Kong, China

[4] Department of Material Science and Engineering, City University of Hong Kong, Kowloon, Hong Kong, China.

[5] State Key Laboratory of Optoelectronic Materials and Technologies, Guangdong Province Key Laboratory of Display Material and Technology, School of Electronics and Information Technology, Sun Yat-sen University, Guangzhou, China

[6] Dongguan Institute of Opto-Electronics, Peking University, Dongguan, China

[7] School of Microelectronics, Southern University of Science and Technology, Shenzhen, China

[8] Department of Applied Physics, Research Institute for Smart Energy, The Hong Kong Polytechnic University, Hung Hom, Hong Kong, China

[9] School of Biomedical Sciences, The University of Hong Kong, Hong Kong, China

‡ These authors contributed equally to this work.

* E-mail: Y.L. (ylin@hku.hk) and Z.Q.C. (zqchu@eee.hku.hk).

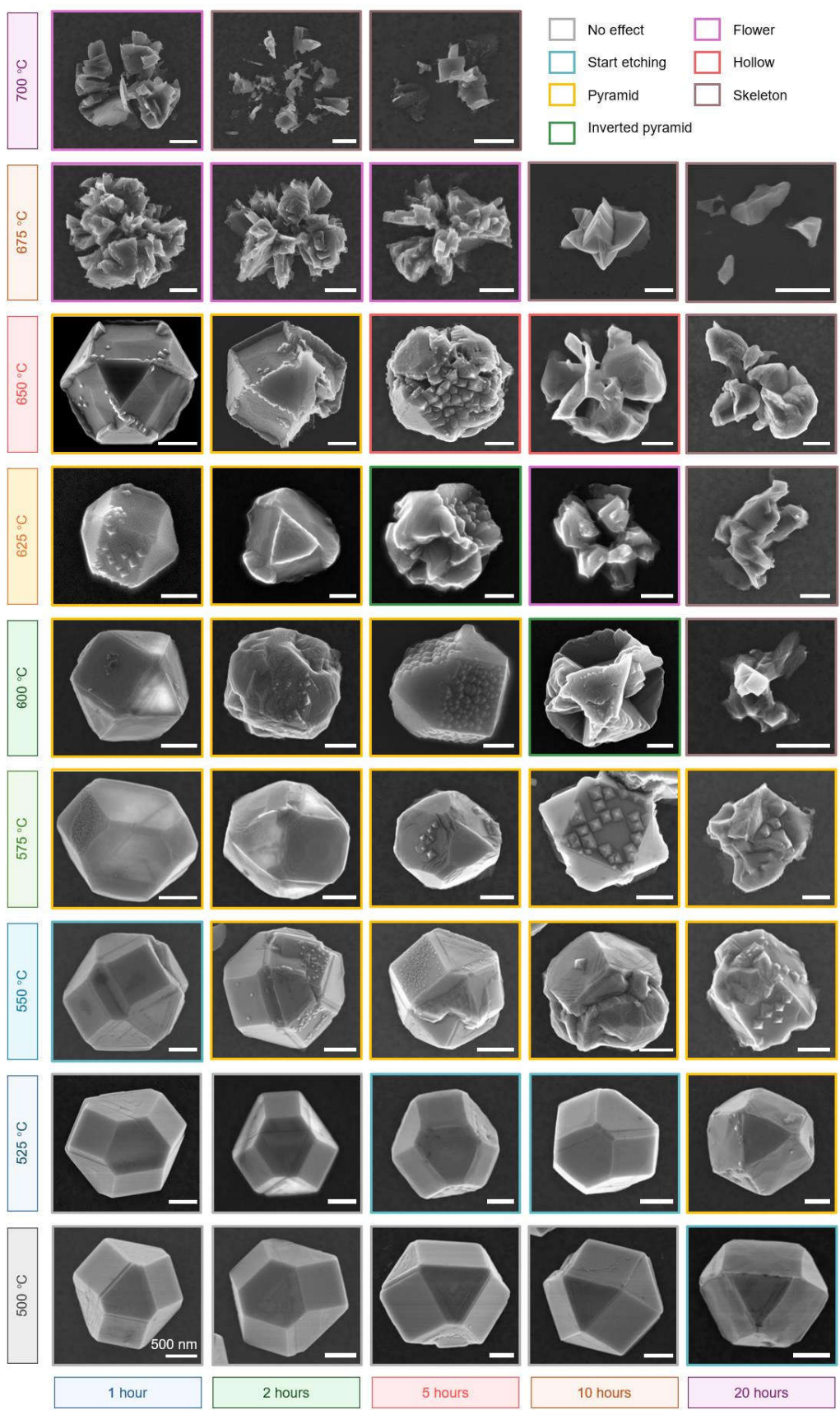

**Figure S1.** The SEM images of the diamond particles after different nanosculpting conditions. Temperature: from 500 to 700°C, duration: from 1 to 20 hours. Different shapes are achieved under different conditions.

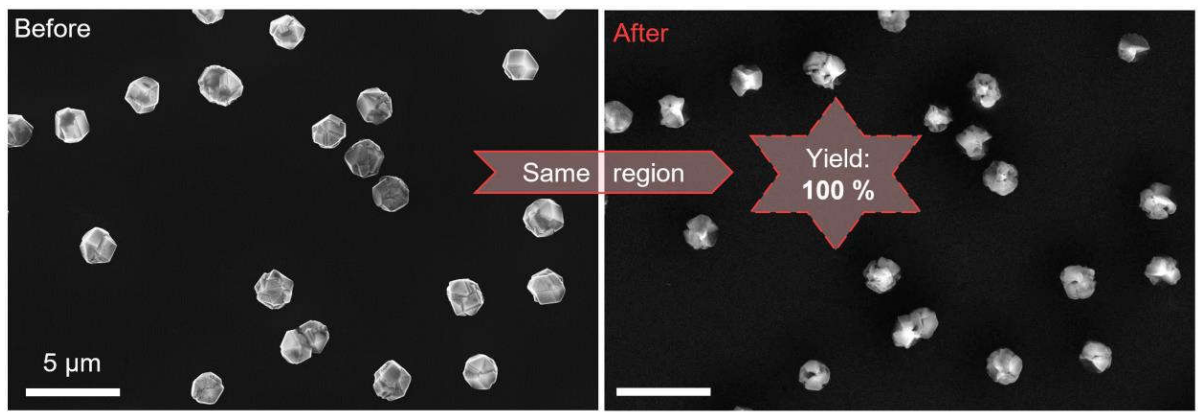

**Figure S2.** 100% yield. SEM images of the same region before and after nanosculpting process, with 100% yield of nanosculpting the particles.

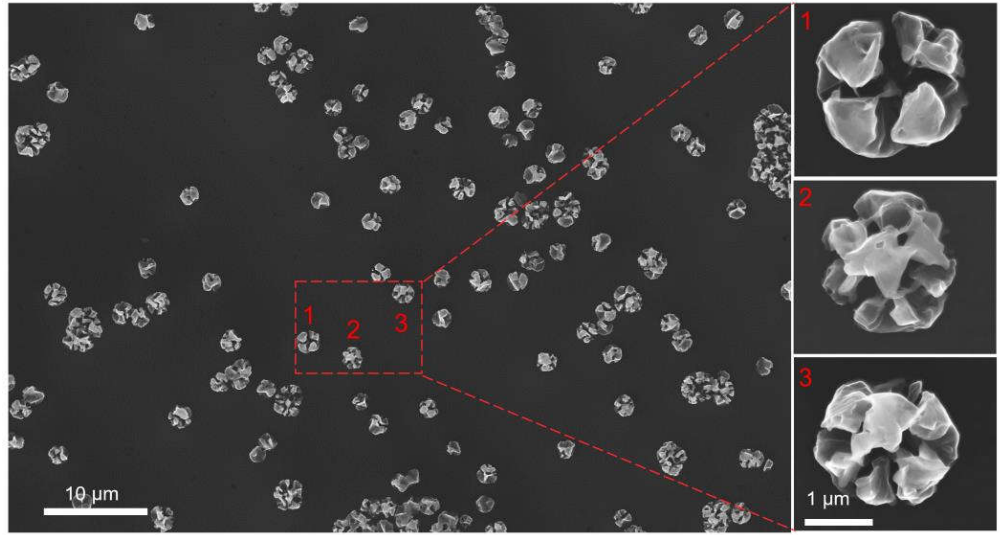

**Figure S3.** Uniformity. SEM image of the reshaped hollow diamond sample shown in a large field of view, with all the diamond particles changed into hollow shape, indicating the uniformity of the nanosculpting process to achieve the desire shape.

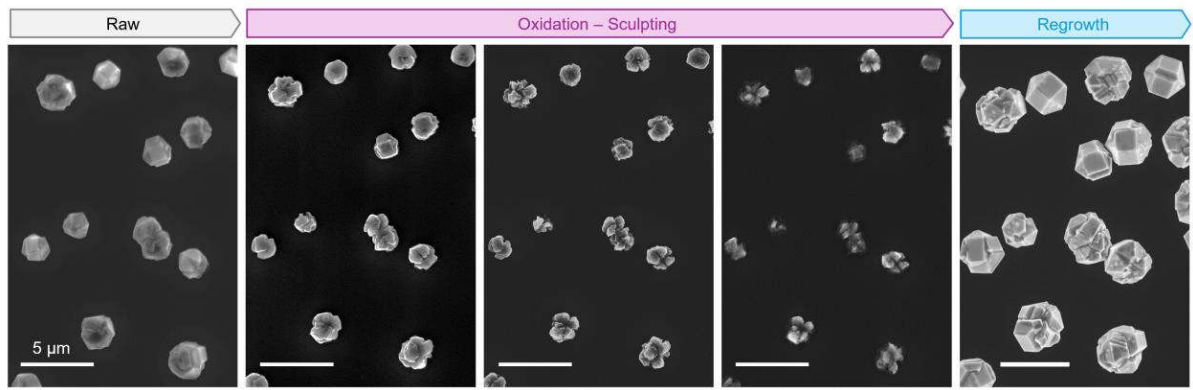

**Figure S4.** Regrowth of diamond using the reshaped diamond particles as seeds. When the reshaped diamond particles are used as seeds for diamond growth, the shapes of the newly grown particles can revert to their original shape.

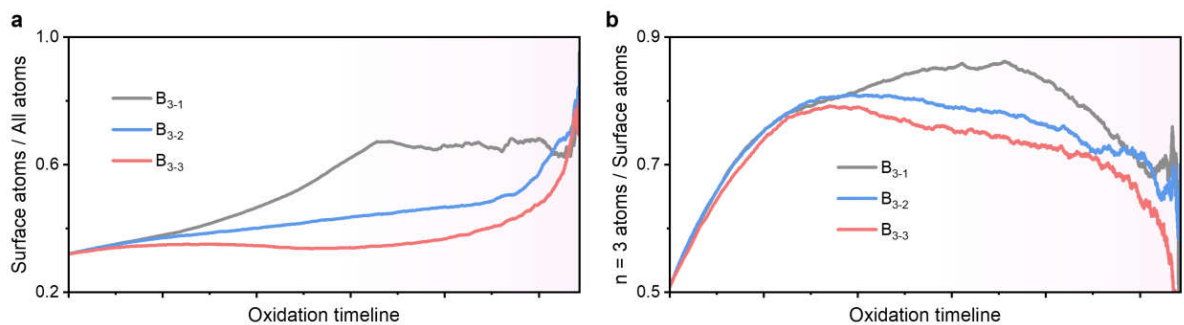

**Figure S5.** Effect of oxidation conditions on atomic ratios of the particle. Ratios of (a) the amount of surface atoms to the total amount of atoms and (b) the amount of $n = 3$ atoms to the surface atoms as a function of oxidation time for the same particle at different oxidation conditions from low to high levels (*i.e.*, $B_{3-1}$ to $B_{3-3}$ in Figure 6 of the main text), indicating the transition of etching modes (from outside-in to inside-out) by changing the oxidation conditions.

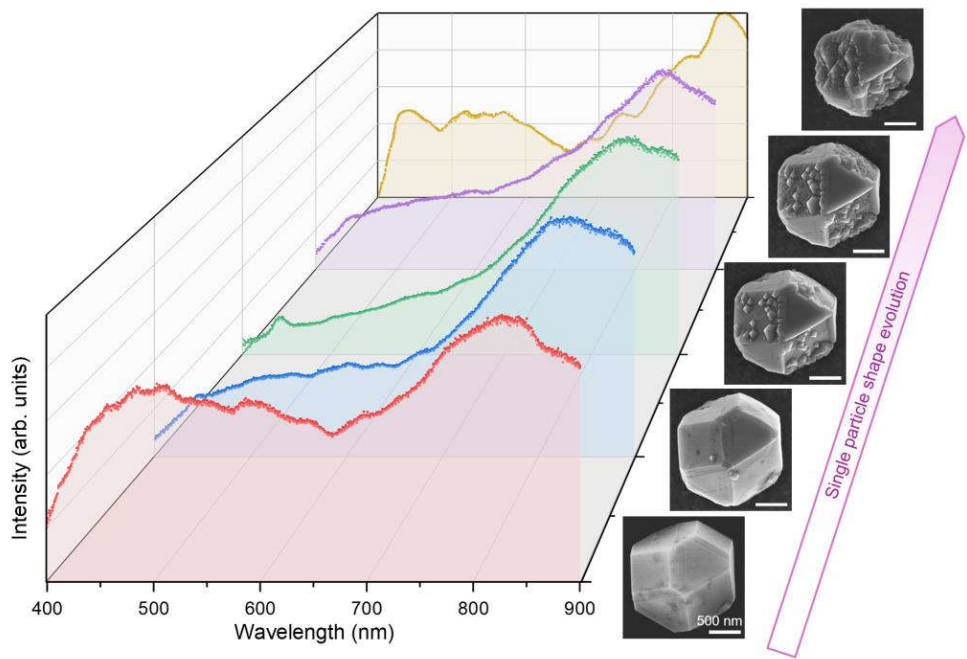

**Figure S6.** Measuring shape-dependent optical properties at individual particle level. The evolution of scattering spectrum of single particle upon air oxidation nanosculpting. The corresponding SEM images of the measured diamond particle are shown next to the scattering spectrum.